%% file: main.tex
\documentclass{sig-alternate-10pt}
\usepackage{epsfig}
\usepackage{amsmath}
\usepackage{subfigure}
\usepackage{color}
\usepackage{multirow}
\usepackage{graphicx}

\usepackage[T1]{fontenc}
\usepackage[utf8]{inputenc}

\usepackage{mathtools}
\usepackage{hyphenat}

\begin{document}
\title{Adaptive and application dependent runtime guided hardware prefetcher reconfiguration on the IBM POWER7}

\numberofauthors{1}
\author{
	David Prat, Cristobal Ortega, Marc Casas, Miquel Moret\'{o}, Mateo Valero  \\ \\
  \affaddr{Barcelona Supercomputing Center (BSC)} \\ 
	\affaddr{Universitat Polit\`ecnica de Catalunya}\\
	\affaddr{08034, Barcelona, Spain}\\
}

\maketitle
\begin{abstract}
\input{00-abstract}
\end{abstract}

\section{Introduction}
\label{sec:intro}
\input{01-introduction}

\section{Background}
\label{sec:Background}
\input{02-background}

\section{Adaptive prefetcher}
\label{sec:adaptiveprefetcher}
\input{03-adaptiveprefetcher}

\section{Results evaluation}
\label{sec:resultsevaluation}
\input{04-resultsevaluation}

\section{Related work}
\label{sec:relatedwork}
\input{05-relatedwork}

\section{Conclusions}
\label{sec:conclusions}
\input{06-conclusions}


\section*{Acknowledgements}
This work has been partially supported by the Spanish Ministry of Science and Innovation under grant TIN2012-34557, the HiPEAC Network of Excellence, by the European Research Council under the European Union's 7th FP, ERC Grant Agreement n. 321253, and by a joint study agreement between IBM and BSC (number W1361154). Miquel Moreto has been partially supported by the Ministry of Economy and Competitiveness under Juan de la Cierva postdoctoral fellowship number JCI-2012-15047. With the support of the Secretary for Universities and Research of the Ministry
of Economy and Knowledge of the Government of Catalonia and the Cofund programme of the Marie Curie
Actions of the 7th R\&D Framework Programme of the European Union (Contract 2013 BP\_B 00243)

\bibliographystyle{abbrv}
\bibliography{biblio}
\end{document}

%% file: 00-abstract.tex
Hardware data prefetcher engines have been extensively used to reduce the impact of memory latency. However, microprocessors' hardware prefetcher engines do not include any automatic hardware control able to dynamically tune their operation. This lacking architectural feature causes systems to operate with prefetchers in a fixed configuration, which in many cases harms performance and energy consumption.

In this paper, a piece of software that solves the discussed problem in the context of the IBM POWER7 microprocessor is presented. The proposed solution involves using the runtime software as a bridge that is able to characterize user applications' workload and dynamically reconfigure the prefetcher engine. The proposed mechanisms has been deployed over OmpSs, a state-of-the-art task-based programming model. The paper shows significant performance improvements over a representative set of microbenchmarks and High Performance Computing (HPC) applications.


%% file: 01-introduction.tex
Hardware data prefetch is a performance optimization technique that helps to alleviate the so-called \emph{Memory Wall}~\cite{Wulf.1995} problem by taking advantage of applications' spatial locality when accessing to memory. Although some contemporary processors come with a set of knobs that adjust different parameters of the hardware prefetcher, their tuning is left to programmer's responsibility being them set to a default configuration when the  system boots up. Unfortunately, apart from being a source of detriment for application's performance, in some cases, this default configuration can suppose a waste of consumed power. For example: prefetching a great amount of data in each memory request may involve bringing unnecessary data that not only wastes power by overloading memory bandwidth, but also pollutes cache memory hierarchy potentially reducing the effective cache space, which can impact performance in multicore environments.

The IBM POWER7 microprocessor~\cite{kalla2010power7} provides the user with the possibility to enable/disable the hardware prefetcher, also to tune the depth of each prefetch operation, to find store prefetch streams of data and to find strides in data accesses, which are gaps of a given fixed size in a data stream. Over this paper, it will be shown how different applications can benefit from this sort of knobs and it will be made evident that hardware prefetcher configuration can not be left to randomness nor default values but it needs of an algorithm that finds a balance in power-performance depending on each application workload. To provide the algorithm with data to determine which configuration to choose, placed in the runtime, a dynamic mechanism that can track performance of multithreaded workloads will be constructed. Specifically, the dynamic mechanism collects performance counters at task level thus being possible to adjust the prefetcher configuration for each code region delimited by the programmer.

This paper is organized as follows: Section 2 describes the IBM POWER7 main characteristics and the prefetcher reconfigurability. Section 3 describes the proposed dynamic mechanism that finds the best prefetcher configuration at runtime. Next, Section 4 consists in an evaluation of the proposed solutions by means of analyzing performance metrics of selected representative benchmarks. Section 5 summarizes the related work and, finally, Section 6 presents the conclusions of this paper.

%% file: 02-background.tex
The IBM POWER7~\cite{kalla2010power7} is an 8-way issue superscalar symmetric multiprocessor based on the Power Architecture. Its main specifications include: 8 cores with 4-way SMT; for each core, two separated L1 caches of 32KB, one for data and other for instructions, plus a 256KB L2 cache. Furthermore, there is an on-chip 32MB L3 shared cache where each core has its private 4MB portion, being able to access other portions though at a cost of higher latency.
The IBM POWER7 reconfigurability allows the end-user to choose the SMT degree, it can be set to single-thread, two-way and four-way. There is also the possibility to change the priority in the decoded instructions of each thread and there are also different knobs associated to the hardware data prefetcher that control its operation mode.


\begin{table}[!t]
	\begin{center}
	\caption{Hardware prefetcher configurations}%
	\label{table:prefetchertable}
	\footnotesize
		\begin{tabular}{ | l | l || l | l |}
		\hline
		DSCR & Description & DSCR & Description \\ \hline
		xx001 & Off (disabled) & xx101 & Deep \\
		xx000 & Default (Deep) & xx110 & Deeper \\
		xx010 & Shallowest & xx111 & Deepest \\
		xx011 & Shallow & x1xxx & Prefetch on stores \\
		xx100 & Medium & 1xxxx & Stride-N \\
		\hline
		\end{tabular}
	\end{center}
	\vspace{-0.3cm}
\end{table}

The IBM POWER7's hardware data prefetcher is programmable per each SMT hardware thread, which means that there are 32 configuration registers accessible from the Operating System (OS). They are denoted as Data Stream Control Register (DSCR). They operate independently, which means that while one hardware thread is executing aggressively prefetching data, another one can be running with the prefetcher disabled.
It is possible to enable or disable the prefetcher engine in each thread as well as to change the depth of each prefetcher operation. 
Moreover, detecting store data streams and strided accesses can also be enabled. Table \ref{table:prefetchertable} shows how to do it by writing the DSCR.
Bits first to third are called default prefetcher depth (DPFD) where their value represent, in each case, the number of lines each prefetch operation brings from main memory to cache. The fifth bit, called Stride-N Stream Enable (SNSE), only has some effect in its activation if the fourth bit, called Store Stream Enable (SSE), is also enabled and the hardware data prefetcher is enabled. 

In this paper, OmpSs~\cite{DuranABLMMP11}, a state-of-the-art task-based programming model is used. This programming model, similarly to the recent OpenMP 4.0 standard, lets the programmer to specify sequential regions of code with their data dependencies. These code regions are called tasks and can run once their input and control dependencies are satisfied. The OmpSs runtime system orchestrates the parallel execution of the different tasks while makes sure all the dependences are satisfied.

%% file: 03-adaptiveprefetcher.tex

In this paper, an adaptive prefetcher mechanism able to operate at runtime is proposed. Performance metrics associated to application's execution will be used to choose the most suitable configuration.
The mechanism operates in two phases: During the \emph{exploration} phase, each prefetcher configuration is evaluated in terms of performance improvement. During the \emph{stable} phase, the best prefetcher configuration found in the exploration phase is used for another amount of consecutive tasks.
A very similar approach about hardware prefetcher reconfiguration at runtime was recently presented by Jimenez et al.~\cite{victor.2012}. In their work, a fixed time of 10ms for the exploration phases as well as 100ms for the stable phases were proposed. These values were chosen empirically and aimed to mitigate the problem that appears when one prefetcher configuration is chosen but the application phase changes to another one that can benefit more from a different prefetcher configuration. However, their mechanism selects the same prefetcher configuration for all threads of an application in the stable phase. In this paper, a more powerful technique is presented, using granularity at OmpSs task level to characterize these phases and allowing to have different prefetcher configurations per task type, even if they run simultaneously.


The lengths of exploration and stable phases are important parameters of the adaptive mechanism. These lengths are measured in terms of number of task instances that are executed in the corresponding phase. 
The execution time of each task instance is measured and stored internally in the runtime system metadata. Regarding exploration phases, it is necessary to run enough experiments to filter the measurement noise while keeping the exploration phase as short as possible. First experiments are about finding optimal values for exploration and stable phases. 
Section~\ref{subsec:impactofpl} explains in detail the exploration we do in this paper regarding exploration phases' lengths.

The impact of having different optimal prefetcher configurations for different task types instead of having a task agnostic mechanism is also evaluated in Section~\ref{subsec:discerningtt}.
Different OmpSs tasks may have different kinds of workloads and therefore they can benefit more from different prefetcher configurations. However, that difference may not be large enough to compensate the additional overhead that this task type aware mechanism has.

The trade-offs between performance improvement and power consumption in terms of memory bandwidth usage are explored in Section~\ref{subsec:ipcdrivenppo}.
The paper presents and evaluates a solution based on an \(\epsilon\) parameter configurable by the user to determine what percentage of difference in the IPCs of one prefetcher configuration with respect to another one less aggressive is needed to choose it as optimal. 
In this case, it is important to see for each application the relation between the aggressiveness of the prefetcher configuration, the used bandwidth and the execution time. When the adaptive prefetcher mechanism chooses aggressive prefetcher configurations, higher bandwidth rates are consumed. 
However, consuming more bandwidth with small performance improvement may not be worth.

%% file: 04-resultsevaluation.tex
\subsection{Experiments Setup}
\label{subsec:expsetup}
In this work the used system has been an IBM BladeCenter PS701; which basically is a blade containing one socket with an 8 core IBM POWER7 running at 3.0~GHz. Although the POWER7 has two quad-channel memory controllers, the PS701 uses a single memory controller offering up to 40GB/s of bandwidth. The OS is SUSE Linux Enterprise Server 11 SP3. Applications have been compiled with Mercurium 1.99.1 source-to-source compiler using as back-end compilers IBM XL C/C++ 11.1 and IBM XL Fortran 13.1. Prefetcher instructions added by the back-end compiler have been disabled as only the dynamic mechanism within the runtime will be in charge of configuring the prefetcher. Prefetcher configurations' performance monitoring has been done by collecting hardware counters using PAPI. Because bandwidth results involve dealing with shared performance counters, perf, the original implementation from Linux kernel, has been used.

Regarding the used benchmarks, applications written in OmpSs programming model have been chosen from different sources as well as own written codes aiming to do stress tests in the system. Benchmarks are briefly explained here.

\begin{itemize}
	\item \textbf{Dotproduct} (DP): an OmpSs implementation of a dot product of two vectors $a$ and $b$ with a stride $K$ ($DP_K=\sum_i a[K\cdot i]\cdot b[K\cdot i]$). This microbenchmark was specifically created to test the hardware prefetcher in a controlled environment. 
	\item	\textbf{Jacobi}: it computes the solution of a linear system obtained from a stencil scheme via the Jacobi iterative method.
	\item	\textbf{K-means}: it performs K-means clusterings, that is, partitions n observations into k clusters in which each observation belongs to the cluster with the nearest mean.
        \item	\textbf{Knn}: an implementation of a machine learning non-parametric method used for classification and regression called k-nearest neighbors.
	\item	\textbf{Specfem3D}: this application simulates a 3D seismic wave propagation in any region of the Earth based on the spectral-element method~\cite{Patera.1984}.
	\item \textbf{Heat}: it solves linear systems that come from heat distribution problems. 
\end{itemize}

By means of the Dotproduct benchmark, we validate the expected behavior of the IBM POWER7 prefetcher. With a linear access pattern ($K=1$), enabling the prefetcher halves execution time. When the stride equals the cache line size, the aggressiveness of the prefetcher is critical, obtaining 5x speedups with the deepest prefetcher with respect to disabling it. When the stride is larger than twice the size of the cache line, the SNSE bit has to be set to observe performance improvements. Finally, these benchmarks are not sensitive to the SSE bit, since they accumulate the result in a single variable. Instead, if we compute the addition of two vectors and store it in a target vector, then the SSE bit significantly improves performance when activated. In the remaining of the paper, we assume $K=1$ for the Dotproduct benchmark.

\subsection{Impact of Phases Lengths}
\label{subsec:impactofpl}
The first step to deploy a successful adaptive technique is to evaluate the impact of phases lengths and figure out their optimal values. 
Exploration phases have to be wide enough to make sure that the phase is representative and thus the optimal prefetcher configurations can be extrapolated.

We tested Dotprod, Jacobi, Spefem3D and Heat with 1 and 8 threads computing for each case the relative IPC error of setting the prefetcher beforehand with respect to use the dynamic reconfiguration. So the idea is, for different lengths of exploration phases, to compute the relative error of the IPC in exploration phases with respect to executions when setting the prefetcher beforehand.

In general, results for lengths of 2, 4, 8, 16, 32 and 2500 tasks in each prefetcher configuration did not show a significant improvement in the relative error but, few applications showed a sensitive drop in the error when using lengths starting at 8 and 16 tasks. Regarding the biggest length, while sometimes benefiting from it, the error suffered from a high increment in general; that is because many OmpSs applications do not have so many task instances for some task types thus it is not possible to try all prefetcher configurations.

We did not observe a high correlation between execution times and different orders of magnitude in lengths of exploration phases. For this we decided to choose a length that allowed most of OmpSs applications to execute several times exploration phases. 

\begin{figure*}[bht!]
	  \center
		\includegraphics[ scale=0.99, width=\textwidth] {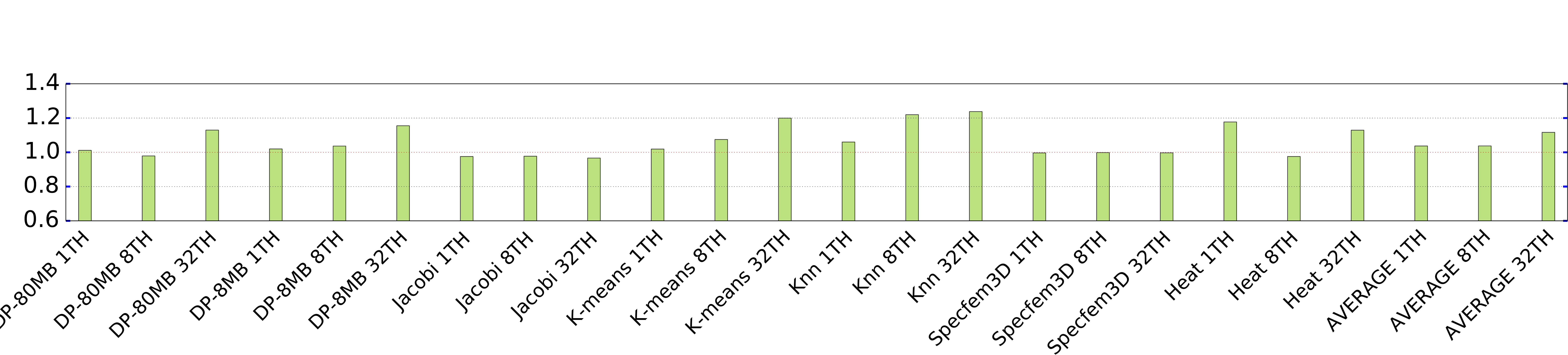}
		\vspace{-0.3cm}
		\caption{Performance speedup when classifying statistics per task type}%
		\label{fig:sl}%
\end{figure*}

\subsection{Impact of Classifying Task Types}
\label{subsec:discerningtt}
Next experiment consists in comparing performance of two versions of the dynamic mechanism: The first one classifies different task types when choosing the best prefetcher configuration, which implies gathering statistics for task types separately. The second approach treats all task types in the same way.

Separating different task types can give better results because some types may benefit more from a given prefetcher configuration whereas other task types may benefit more from a different configuration. 
However, this difference could not be enough to compensate an additional complexity of dealing with task types separately.

Figure \ref{fig:sl} shows results of this experiment considering different applications and different requested thread numbers. 
Speedups are calculated with respect to the version that does not classify by task type.
Additionally, when we apply the task type aware approach we deploy an additional optimization that consists in saving the prefetcher configuration per thread in the runtime meta-data, which avoids consulting and modifying the prefecther status very often. 
The Dotproduct benchmark only has one task type, so the speedup observed when considering the 32 threads executions is obtained from this additional configuration. 
We included Dotproduct in this set of experiments to evaluate these extra benefits, which are orthogonal to the task type aware approach.
The rest of applications, which have more than one kind of task, show different behaviors. 
While Jacobi and Specfem3D do not present speedups, Kmeans, Knn and Heat have speedups starting at parallel level of 1 thread. 
The classifying version of the dynamic mechanism is considered as a useful improvement since it provides significant performance benefits.

\subsection{IPC Driven Power-Performance Optimization}
\label{subsec:ipcdrivenppo}
Having determined acceptable lengths for exploration and stable phases for any OmpSs application and with a dynamic mechanism classifying performance statistics of tasks by their type, the third experiment consists in saving power by reducing the aggressiveness of the prefetcher when this one does not bring a considerable gain in performance. 
This is done through a configurable parameter \(\epsilon\) that represents, in terms of percentages, the difference in the IPC of one aggressive configuration with respect to another one less aggressive. 
When the difference is smaller than the \(\epsilon\), the most aggressive configuration is not considered to be better.
The method starts from the disabled prefetcher configuration, and goes through more aggressive configurations step by step. 

\begin{figure*}[bht!]
          \center
                \includegraphics[ width=\textwidth ] {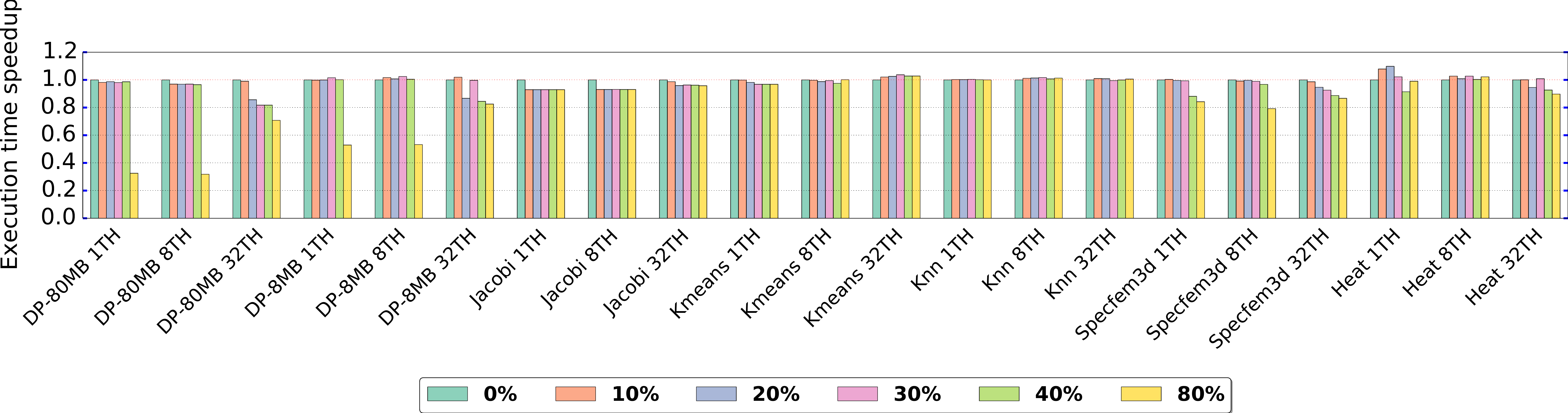}
                \includegraphics[ width=\textwidth ] {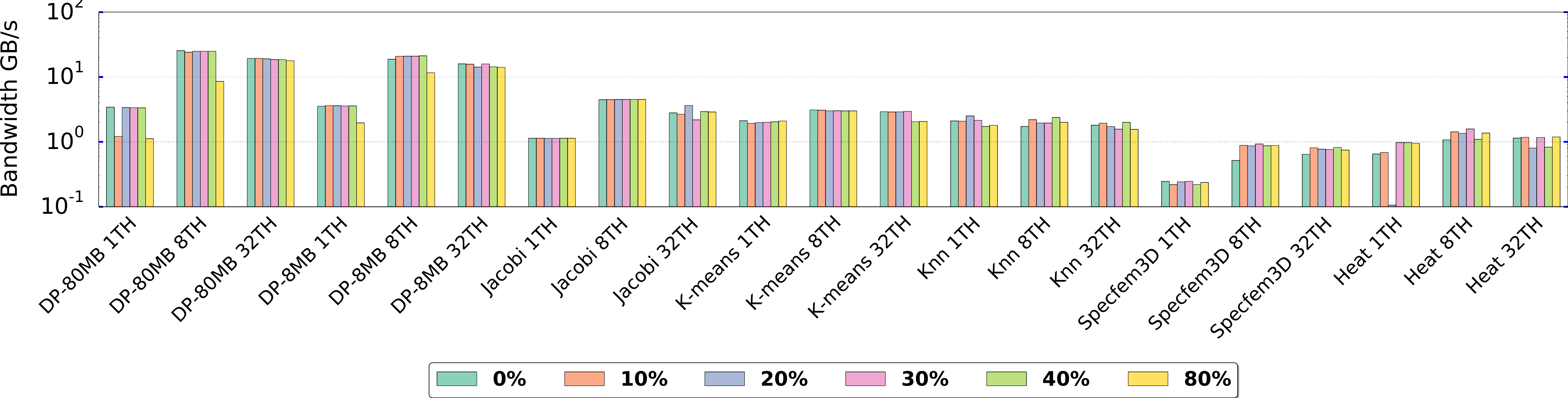}
                \caption{Performance speedup and bandwidth when \(\epsilon\) increases}%
                \label{fig:ps}%
\end{figure*}

Figure \ref{fig:ps} shows execution slowdowns considering \(\epsilon\) values of 0, 10, 20, 30, 40 and 80\%.
Dotproduct application results show a drop in the performance when \(\epsilon\) sets an 80\% of difference in the IPC. 
This 80\% of difference in the IPC is observed when passing from prefetcher disabled to enabled in the shallowest configuration, which is the first step of increasing prefetcher aggressiveness. 
Jacobi presents nearly a 10\% of slowdown for 1 and 8 threads configurations when setting \(\epsilon\) to 10\%. 
Specfem3D interestingly shows a drop in the performance nearly in each step that heightens \(\epsilon\), showing a strong correlation between the depth of the prefetcher and the obtained IPC.

Regarding the memory bandwidth usage, both Dotproduct and Specfem3D show a reduction that is consistent with the performance drop, meaning in these cases the applications fully exploit the extra bandwidth used by the prefetchet. 
Jacobi, knn and Heat do not show a consistent reduction in the used bandwidth and they neither suffer performance slowdowns when choosing less aggressive configurations.  
Finally, K-means application does not suffer from performance slowdowns in the execution time although the bandwidth usage gets significantly reduced when \(\epsilon\) increases. 
Therefore, in this particular application, the adaptive mechanism successfully selected the less aggressive prefecther configuration that provides maximum performance, avoiding the spending of useless memory bandwidth.

%% file: 05-relatedwork.tex
There have been many works that have dealt with data prefetch \cite{Baer.1991, Jouppi.1990, Palacharla.1994}. First attempts were based on sequential prefetchers, this approach suggests to prefetch memory blocks sequentially. Despite being effective in these cases, this solution is not able to yield performance when the application does not follow a sequential data access pattern. Due to this, further research in prefetchers was done to try to capture the non-sequential nature of those applications. Prefetch techniques aimed to deal with pointer-based applications have been studied~\cite{Ebrahimi.2009, Roth.1998, Yang.2000}. Solihin et al.~\cite{Solihin.2002} made use of a user-level memory thread to do prefetching, getting in the applications with irregular accesses significant speedups. Joseph and Grundwald~\cite{Joseph.1997} worked on Markov-based prefetchers. Although most of these works about prefetching have not been put into practice with real processors, limit studies and prefetch analytical models have been proposed~\cite{Emma.2005, Srinivasan.2004}.

A further step in data prefetching is to consider the interaction between threads that take place in the CMP processors. Ebrahimi et al. \cite{Ebrahimi.2011} and Lee et al. \cite{Lee.2008} study the effect of thread-interaction on prefetch and design prefetch systems that improve throughput and fairness. Liu and Sohilin~\cite{Liu.2011} present a study about the impact prefetching has and bandwidth partitioning in CMPs.
Although there are many sutudies about data prefetching on top of simulators, there are very few works that make use of real processors. For instance, Wu and Martonosi~\cite{wu2011ispass}	 characterize the prefetcher of an Intel Nehalem processor and provide a straightforward algorithm that can control dynamically the activation and deactivation of the prefetcher. Nevertheless, their work only contemplates intra-application cache interference obviating actual system performance. Liao et al.~\cite{Liao.2009} build a machine learning model that dynamically modifies the prefetch configuration of the machines in a data center (based on Intel Core2 processors). Their work also bases its approach on turning on and off the prefetcher.

Beyond enabling and disabling the prefetcher, there are other kind of works targeted to control thread execution rate. For example, playing with fetch policies within a SMT processor has been studied~\cite{Cazorla.2006, Choi.2006}. They aim to increase throughput and/or provide quality of service (QoS). In the same line, the work of Boneti et al.~\cite{Boneti.2008} study the usage of the dynamic hardware priorities in the IBM POWER5 processor aiming to yield performance from resource balancing and prioritization. Qureshi and Patt \cite{Qureshi.2006} study how to improve throughput through solving the problem of partitioning the last-level cache for multiple applications. Moret\'{o} et al. \cite{Moreto.2009} show a similar solution based on achieving QoS for multiple applications running at the same time.

%% file: 06-conclusions.tex
Contemporary microprocessors are being designed with reconfigurability features and increasingly more capable of counting different events by means of hardware counters. In this paper, a portable solution implemented within a runtime smartly reconfigures the hardware prefetcher making use of hardware counters. A dynamic mechanism makes the process of reconfiguration automatic. Once it has enough collected performance data from different configurations, it calls to an algorithm that is in charge of determining which one is the most power-performance efficient. This process is repeated with a given timing during the application execution. A series of experiments have shown that sensitivity in performance is nearly negligible when collecting great amounts of data from performance counters; this can be attributed to the fact that few OmpSs tasks contain performance data that turns out to be representative enough. Additionally, OmpSs task types classification has a positive impact in performance because different OmpSs task types may benefit from different prefetcher configurations as task types may determine different kinds of workloads in the machine. Finally, a proposal for saving power in the cases in which aggressive prefetcher configurations do not come with a substantial speedup has proved to be potentially useful and reaped good results. The underlying idea is to set an IPC percentage threshold that limits the aggressiveness of chosen prefetcher configurations. 